\documentclass[prd,fleqn,twoside,twocolumn]{revtex4}

\usepackage{latexsym,exscale}
\usepackage{epsfig}
\usepackage{amsmath,amssymb,amsfonts}

\newcommand{\bra}[1]{\left\langle #1 \right|}  
\newcommand{\ket}[1]{\left| #1 \right\rangle}
\newcommand{\sla}[1]{#1 \hspace{-1.1ex}\slash\hspace{0.3ex}}
\newcommand{\Sla}[1]{#1 \hspace{-1.6ex}\slash\hspace{0.7ex}}
\DeclareMathOperator{\tr}{Tr}

\begin{document}

\title{Resummation of nuclear enhanced higher twist in the Drell Yan process}
\author{Rainer J.~Fries}
\affiliation{Department of Physics, Duke University, P.O.Box 90305,
  Durham, NC 27708}
\affiliation{Institute for Theoretical Physics, University of Regensburg, 
  93040 Regensburg, Germany}
\begin{abstract}
We investigate higher twist contributions to the transverse momentum broadening of
Drell Yan pairs in proton nucleus collisions. We revisit the contribution of matrix elements
of twist-4 and generalize this to matrix elements of arbitrary twist. An estimate of the 
maximal nuclear broadening effect is derived. A model for nuclear enhanced matrix elements
of arbitrary twist allows us to give the result of a resummation of all twists
in closed form. Subleading corrections to the maximal broadening are discussed qualitatively.
\end{abstract}

\maketitle

\section{Introduction}

In the era of RHIC and LHC the understanding of hard processes in heavy ion 
collisions plays an increasingly important role. It has become clear
that the onset of the regime of perturbative quantum chromodynamics (pQCD) 
with increasing momentum transfer in reactions involving large nuclei can 
be very different from that in collisions of single hadrons. 
Previous studies have shown that the leading twist approximation
\cite{CSS:mueller} in pQCD, even at intermediate momentum transfers, can be
spoiled by higher components of the twist expansion. 

The reason for that is that matrix elements of higher twist can be increasingly sensitive 
to the size of the nucleus. With naive power counting, twist-2 
parton distributions scale with the mass number $A$ of the nucleus, 
while matrix elements of twist-$(2m+2)$, $m \in \mathbb{N}$, can scale with $m$ additional
powers of $A^{1/3}$. These matrix elements are then called
nuclear enhanced. They correspond to processes with multiple scattering.
For large nuclei these additional factors of $A^{1/3}$ can compensate the
inherent power suppression of higher twist.

Let us explore the physical picture \cite{RJF:hir} behind this for the example 
of $p+A$ collisions.
In the leading twist approximation only one parton $a$ from the nucleus
and one parton $b$ from the proton participate in 
the hard scattering $a+b$. They are described by parton distributions respectively. 
A nuclear enhanced twist-4 matrix element for the nucleus 
corresponds to two partons $a$ and $a'$ entering the hard scattering, 
They enable us to describe the process where the parton from the proton scatters off these
two partons: $a\, a'+b$.
The additional factor of $A^{1/3}$ arises when the
two partons come from different nucleons inside the nucleus.
This mechanism was first pointed out by Luo, Qiu and Sterman some years ago 
\cite{LQS:92,LQS:94sv,LQS:94} based on generalized factorization theorems \cite{QS:91}. 

In the following years various calculations on the level of nuclear enhanced
twist-4 (double scattering) became available for the Drell Yan process $p+A\to l^+ l^-$ in 
proton nucleus collisions \cite{Guo:98ht,FSSM:99,FSSM:00}. However going beyond the level
of twist-4 remained a technically difficult point. It is known that for
the Drell Yan process in symmetric colliding systems the QCD factorization fails at the level
of twist-6 \cite{DFT}. However in $p+A$ we will formally work in the limit of very large 
$A$
and therefore stay in the leading twist approximation for the proton. This allows us to 
go to arbitrary nuclear enhanced twist for the nucleus \cite{Qiu:hpc}. Nevertheless
a systematic calculation of contributions of arbitrary twist was never done in this
framework.

In this publication we study the transverse spectrum of Drell Yan pairs 
and its nuclear broadening. This was already done on the level of twist-4 
\cite{Guo:98jb}. We will repeat this calculation both in light cone and covariant gauge 
for the strong interaction. In particular we comment on the question of electromagnetic 
gauge invariance. 
After that we generalize the result by calculating the contributions
of operators of arbitrary twist. This is feasible for contributions which lead
to maximal broadening, therefore establishing an upper bound on the nuclear effects
on the transverse momentum spectrum. Besides this there are subleading contributions
which we will also discuss shortly.
We then present some arguments how the higher twist matrix elements can be approximated by 
model descriptions in such a way that we are able to give a closed expression for the
sum of all higher twist contributions.

\section{Nuclear enhanced double scattering}

In this section we will go through the calculation of the twist-4 contribution
to the nuclear enhanced Drell Yan process. 
This computation was already done in covariant gauge by X.\ Guo some time ago \cite{Guo:98jb}. 
We confirm her results. We also present an explicit calculation in light cone gauge,
which bears some difficulties for higher twist. We also comment on the problem of
electromagnetic gauge invariance in this calculation.

\subsection{Light cone gauge}

The Drell Yan cross section, integrated over the angular distribution and the rapidity of
the lepton pair, is
\begin{equation}
  \frac{d{\sigma}}{d{Q^2} d{q_\perp^2}} = \sum_a
  \frac{c_a^2 \alpha}{24 \pi^2 SQ^2} \int d y P^{\mu\nu} W_{\mu\nu},
\end{equation}
where $Q$, $q_\perp$ and $y$ are mass, transverse momentum and rapidity of the virtual photon
and $S=(P_1+P_2)^2$. $P_1$ and $P_2$ denote the four momenta of the nucleus (per nucleon) and
the proton in the center of mass frame respectively. 
The sum runs over all quark flavors $a$ 
and $c_a$ is the charge of a given quark flavour relative to the electron charge.
The hadronic tensor is defined as the matrix element of
two electromagnetic currents
\begin{equation}
  W_{\mu\nu} = \int d^4{y} e^{i q \cdot y} \bra{P_1 P_2} j^\mu(0) j^\nu(y)
  \ket{P_1 P_2}
\end{equation}
in the colliding system. After integrating the angular distribution the tensor $P^{\mu\nu}$
reduces to the polarization tensor of the virtual photon
\begin{equation}
  P^{\mu\nu} = -g^{\mu\nu} + \frac{q^\mu q^\nu}{Q^2}
\end{equation}
where $q$ is the four momentum of the virtual photon \cite{FSSM:00}.
Electromagnetic gauge invariance of course requires that 
$q^\mu W_{\mu\nu} = 0 = W_{\mu\nu} q^\nu$.

\begin{figure*}[tb]
  \begin{center}  
  \epsfig{file=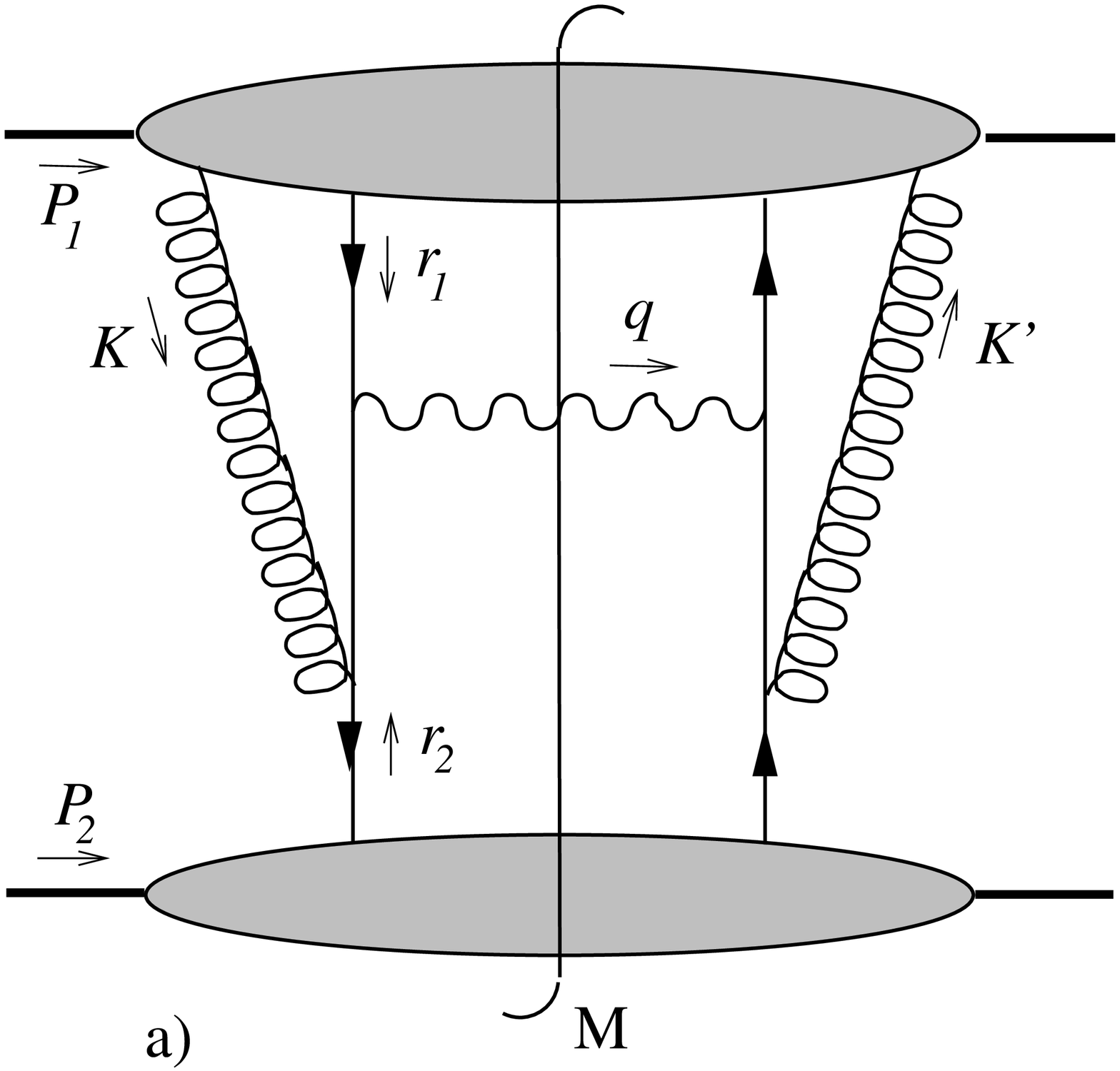,width=4.5cm}\hfil
  \epsfig{file=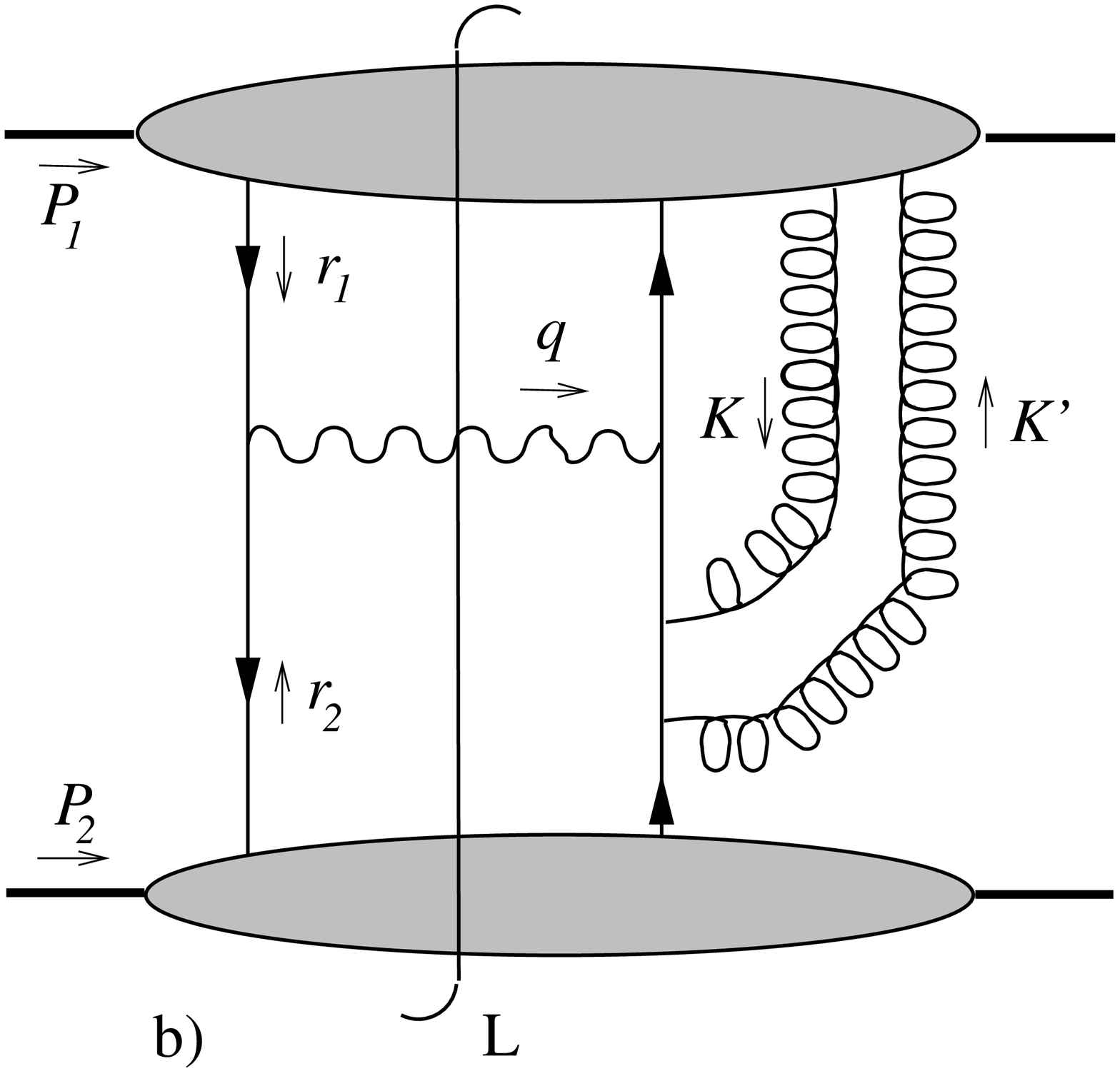,width=4.5cm}\hfil
  \epsfig{file=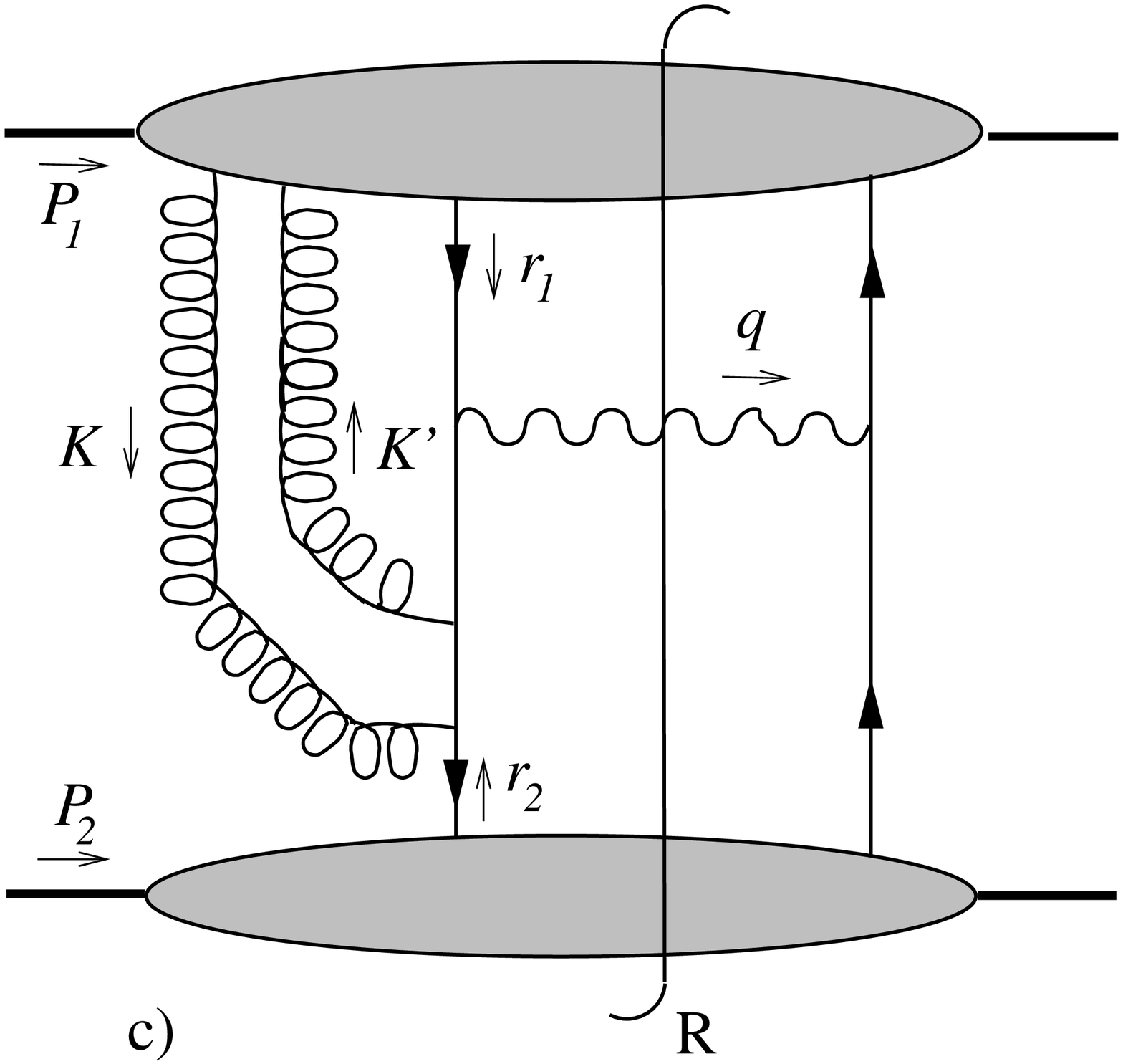,width=4.5cm}\hfil
  \end{center}
  \caption{The diagrams contributing to Drell Yan on the level of twist-4.
  The symmetric diagram (a) is double scattering, the asymmetric diagrams (b) and (c) 
  correspond to an interference of single and triple scattering.}
  \label{fig:dynltlo}
\end{figure*}

We want to calculate the nuclear enhanced twist-4 contribution to the
hadronic tensor. Thence we have to take into account the
diagrams shown in Fig.~\ref{fig:dynltlo}. We denote the momenta of the quarks with $r_1$ and 
$r_2$ and
the momenta of the gluons with $K$ and $K'$. We expand the momenta of the parton lines
between the hard and the soft part into a longitudinal contribution on the light cone
and a transverse contribution
\begin{align}
  r_1^\mu &= \xi_1 P_1^\mu + r_1^\perp & r_2^\mu &= \xi_2 P_2^\mu + r_2^\perp \\
  K^\mu &= x P_1^\mu + K^{\perp \mu} & K^{\prime \mu} &= x' P_1^{\prime\mu}  + 
  K^{\prime \perp \mu}  
\end{align}
introducing parton momentum fractions $x$, $x'$, $\xi_1$ and $\xi_2$.

The hadronic tensor for the symmetric diagram (a) is
\begin{widetext}
\begin{multline} 
  \label{eq:hadtensor}
  W_{\mu\nu}^{(a)} =
  \frac{4 \pi \alpha_s}{2N_c^2} \int d{y} \int 
  \frac{d{\xi_1}}{\xi_1} \frac{d{\xi_2}}{\xi_2}
  \int P_1^+ d{x} P_1^+ d{x'} {d^2}{r_1^\perp} {d^2}{K^\perp} 
  {d^2}{K^{\prime \perp}} f_{\bar q}(\xi_2) 
  (32 \pi^3) \Gamma^{(a)} H^{(a)}_{\mu\nu,\rho\sigma}  
  \frac{1}{(r_2+K)^2 + i \epsilon} \\ \times \frac{1}{(r_2+K')^2 - i \epsilon} 
  \int \frac{d{y_1^-}{d^2}{y_1^\perp}}{(2\pi)^3}  
  \frac{d{y_3^-}{d^2}{y_3^\perp}}{(2\pi)^3}
  \frac{d{y_4^-}{d^2}{y_4^\perp}}{(2\pi)^3} 
  e^{-i\xi_1 P_1^+ y_1^-} e^{-i x P_1^+ y_3^-} e^{i x' P_1^+ y_4^-}
  e^{-i r_1^\perp \cdot y_1^\perp} e^{-i K^\perp \cdot y_3^\perp} 
  e^{i K_1^{\prime \perp} \cdot y_4^\perp} \\
  \frac{1}{2} \bra{P_1} \bar q (0) \gamma^+ q(y_1^-,y_1^\perp)
  A^\rho(y_3^-,y_3^\perp) A^\sigma(y_4^-,y_4^\perp) \ket{P_1}
\end{multline}
\end{widetext}
for the case that the quark comes from the nucleus and the antiquark comes 
from the nucleon. Here 
\begin{multline}
  \Gamma^{(a)} = \delta[Q^2 - (\xi_1+x)\xi_2 S - (r_1^\perp+K^\perp)^2] \\ \times
  \delta[q_\perp^2 + (r_1^\perp+K^\perp)^2] \delta\left[ y- \frac{1}{2} \ln 
  \frac{\xi_1+x}{\xi_2} \right]
\end{multline}
are the momentum conserving $\delta$-functions which connect the photon
momentum to the internal momenta. It is understood that the $\delta$-functions
have to be integrated in order to obtain physical results. We have chosen
to integrate the rapidity dependence of the cross section which is
trivially done by the last $\delta$-function. The first $\delta$-function will be
eliminated by the integral over $\xi_2$.
At a later stage we will also integrate over ${q_\perp^2}$ but since we will do that in 
different ways it is a matter of convenience to keep the unphysical $\delta$-function in the
mean time. $f_{\bar q}$ in above formula is the antiquark distribution in the proton.

The hard part $\cal{H}$ of the cross section is given by the propagators in 
Eq.~(\ref{eq:hadtensor}), the phase factor $\Gamma^{(a)}$ and the trace 
\begin{equation}
  H^{(a)}_{\mu\nu,\rho\sigma} = \frac{1}{4}
   \tr [ \gamma_\nu (-\sla{r}_2 - \Sla{K}') \gamma_\sigma
   \sla{r}_2 \gamma_\rho (-\sla{r}_2 - \Sla{K}) \gamma_\mu \sla{r}_1 ].
\end{equation}
The quark lines connecting the hard part to the nucleus and the nucleon respectively
are contracted
with $\sla{r}_1$ and $\sla{r}_2$ respectively. 
In order to project out the correct twist contribution there is an important step. 
We have to perform a collinear expansion of all momenta entering the hard part 
which is equivalent to a Taylor expansion in all transverse momenta. 
In light cone gauge for the gluon field we can stick to the 0th order of the expansion, 
i.e.\ we simply set all transverse 
momenta equal to zero. In Eq.~(\ref{eq:hadtensor}) this was already done for the quark 
lines from the 
nucleon since we only wanted to generate the ordinary twist-2 parton distribution 
$f_{\bar q}(\xi_2)$. For parton lines from the nucleus we keep the transverse momenta for 
the moment.
The reason is that this collinear limit is not straightforward in light cone gauge.
In fact, it will turn out that some contributions which are falsely taken to vanish 
in the collinear limit in light cone gauge give finite results because the zeros in 
the hard part are canceled by poles elsewhere.
 
The two remaining diagrams (b) and (c) in Fig.~\ref{fig:dynltlo} have the same 
decomposition of the hadronic
tensor but with denominators of the propagators replaced by
\begin{align}
  & \frac{1}{(r_2+K'-K)^2 - i \epsilon} \> \frac{1}{(r_2+K')^2 - i \epsilon} \>, \\
  & \frac{1}{(r_2+K-K')^2 + i \epsilon} \> \frac{1}{(r_2+K)^2 + i \epsilon}
\end{align}
for diagrams (b), (c) respectively, $\delta$-functions
\begin{multline}
  \Gamma^{(b)} = \delta[Q^2 - \xi_1 \xi_2 S - {r_1^\perp}^2] \\ \times
  \delta[q_\perp^2 + {r_1^\perp}^2] \delta\left[ y- \frac{1}{2} \ln \frac{\xi_1}{\xi_2} 
  \right] ,
\end{multline}
\begin{multline}
  \Gamma^{(c)} = \delta[Q^2 - (\xi_1+x-x')\xi_2 S \\[0.5em] 
   - (r_1^\perp+K^\perp-K^{\prime\perp})^2] 
   \delta[q_\perp^2 + (r_1^\perp+K^\perp-K^{\prime \perp})^2 ]  \\[0.5em]  \times
  \delta\left[ y- \frac{1}{2} \ln \frac{\xi_1+x-x'}{\xi_2} \right]
\end{multline}
and traces 
\begin{align}
  H^{(b)}_{\mu\nu,\rho\sigma} &= \frac{1}{4} 
   \tr [ \gamma_\nu (-\sla{r}_2 - \Sla{K}' + \sla{K}) \gamma_\rho \\ \nonumber
   & \qquad \times 
   (-\sla{r}_2 - \Sla{K}') \gamma_\sigma \sla{r}_2  \gamma_\mu \sla{r}_1 ],
  \\
  H^{(c)}_{\mu\nu,\rho\sigma} &=  \frac{1}{4}  \tr [ \gamma_\nu  \sla{r}_2  
  (-\sla{r}_2 - \Sla{K}) \gamma_\rho  \\  \nonumber
   & \qquad \times (-\sla{r}_2 - \Sla{K} + \Sla{K}' ) \gamma_\sigma \gamma_\mu \sla{r}_1 ].
\end{align}

Next we want to convert the gluon gauge fields to field strengths. In light cone gauge $A^+=0$
the transverse components of the $A$ fields give the dominant contribution and we can
use
\begin{multline}
  P_1^+ A^\rho P_1^+ A^\sigma = \frac{1}{ix} (\partial^+ A^\rho) \frac{1}{-ix'}               
  (\partial^+ A^\sigma)     \\  \approx \frac{1}{xx'} F^{+\omega} F_\omega^{\phantom{\omega}+} 
  \left( - g_\perp^{\rho\sigma} /2 \right)
\end{multline}
By the introduction of the field strength tensors we make the poles in the parton momentum 
fractions $x$ and $x'$ explicit that were hidden in the gluon gauge fields. It will turn
out that these poles are artificial and will be canceled by zeros which are
hidden in the hard part. This is a special feature of the calculation in light cone gauge.

We carry out the collinear expansion as the lowest order Taylor expansion of
the hard part in the transverse momenta $r_1^\perp$, $K^\perp$ and $K^{\prime\perp}$
\begin{equation}
  {\cal H}(\perp) = {\cal H}(\perp)\big|_{\perp = 0} + 
   \sum_\perp \perp \frac{\partial}{\partial\perp} {\cal H}(\perp)\big|_{\perp = 0} + {\cal O} (\perp).
\end{equation}
Here $\perp$ shortly stands for all the transverse momenta.
Now we are able to carry out all transverse integrals and to eliminate all dependences on 
transverse coordinates in Eq.~(\ref{eq:hadtensor}) in a trivial way. For diagram (a)
this gives us
\begin{widetext}
\begin{multline}
  \label{eq:hadtensor2}
    W_{\mu\nu}^{(a)} =
  \frac{32 \pi^3 \alpha_s}{N_c^2} \int d{y} \int \frac{d{\xi_1}}{\xi_1} 
  \frac{d{\xi_2}}{\xi_2}
  \int d{x}d{x'} f_{\bar q}(\xi_2) \frac{1}{xx'}
  \left[ {\cal H}(r_1^\perp,K^\perp,K^{\prime \perp})\right]_{\perp \rightarrow 0} 
  \left( - g_\perp^{\rho\sigma} /2 \right)
  \\
  \int \frac{d{y_1^-}}{2\pi}  
  \frac{d{y_3^-}}{2\pi} 
  d{y_4^-}
  e^{-i\xi_1 P_1^+ y_1^-} e^{-i x P_1^+ y_3^-} e^{i x' P_1^+ y_4^-}
  \frac{1}{2} \bra{P_1} \bar q (0) \gamma^+ q(y_1^-)
  F^{+\omega}(y_3^-) F_\omega^{\phantom{\omega}+}(y_4^-) \ket{P_1}.
\end{multline}
\end{widetext}
Due to the subtle interplay of poles and zeros in light cone gauge we evaluate the hard part
with transverse momenta however keeping in mind to take the limit ${\perp \rightarrow 0}$ at
the end. There are some caveats.
E.g.\ if we want to test electromagnetic gauge invariance it is clear that we must not take
$q^\mu \lim_{\perp\rightarrow 0} W_{\mu\nu}$ but we must take 
$\lim_{\perp\rightarrow 0} q^\mu W_{\mu\nu}$ 
since the hadronic tensor contains $\delta$-functions which give $q^\mu$ a finite transverse
component as long as the internal transverse momenta are still finite.

The denominators of the propagators can be transformed into poles in $x$ and $x'$ via
\begin{multline}
  \frac{1}{(r_2+K)^2 + i \epsilon} \> \frac{1}{(r_2+K')^2 - i \epsilon} \\ = (\xi_2 S)^{-2}
  \frac{1}{x+\frac{{K^\perp}^2}{\xi_2 S} + i\epsilon} \> 
  \frac{1}{x'+\frac{{K^{\prime \perp}}^2}{\xi_2 S} - i\epsilon}.
\end{multline}
We define the prefactor and the tensor structure from the field strength into the trace
and finally evaluate
\begin{multline}
  H^{(a)}_{\mu\nu} = (\xi_2 S)^{-2}  H^{(a)}_{\mu\nu,\rho\sigma}   
  \left( - g_\perp^{\rho\sigma} /2 \right)  \\
  = \frac{\xi_1}{\xi_2 S} \Big[ 2 xx' P_{1\> \mu} P_{1\>\nu} + x P_{1\>\mu} 
  K^{\prime\perp}_{\nu} + 
  x' P_{1\>\nu} K^{\perp}_\mu \\ + \frac{1}{2} ( K^{\perp\> \mu} 
    K^{\prime\perp\> \nu} - K^{\perp\> \nu} K^{\prime\perp\> \mu} )
    + \frac{1}{2} K^\perp \cdot K^{\prime\perp} g_{\mu\nu} \\
    - \frac{K^\perp \cdot K^{\prime\perp}}{S} (P_{1\> \mu} P_{2\>\nu} +  P_{2\> \mu} P_{1\>\nu})
    \Big].
\end{multline}
At this stage we see that the naive limit $\perp \rightarrow 0$ carries most of the terms in above
equation to zero. However if we look at it more carefully through the factor $1/xx'$ from the gluon 
fields we will encounter a $0/0$ situation which we have to resolve in order to obtain finite
results.

For diagrams (b), (c) we get similar poles
and the traces reduce to
\begin{multline}  
  H^{(b)}_{\mu\nu} = (\xi_2 S)^{-2}  H^{(b)}_{\mu\nu,\rho\sigma}   
  \left( - g_\perp^{\rho\sigma} /2 \right)  \\
  = \frac{x'\xi_1}{\xi_2 S} \Big[ P_{1\>\nu} (K^{\prime\perp}-K^\perp)_\mu \\ + \xi_2 
  (P_{1\> \mu} P_{2\>\nu} +  P_{2\> \mu} P_{1\>\nu}) - \frac{1}{2} \xi_2 S g_{\mu\nu} \Big] \>, 
\end{multline}
\begin{multline}
  H^{(c)}_{\mu\nu} = (\xi_2 S)^{-2} H^{(c)}_{\mu\nu,\rho\sigma}   
  \left( - g_\perp^{\rho\sigma} /2 \right)  \\
  = \frac{x \xi_1}{\xi_2 S} \Big[ P_{1\>\mu} (K^\perp-K^{\prime\perp})_\nu \\ + \xi_2 
  (P_{1\> \mu} P_{2\>\nu} +  P_{2\> \mu} P_{1\>\nu}) - \frac{1}{2} \xi_2 S g_{\mu\nu} \Big]
\end{multline}
We carry out the integrals in $x$ and $x'$ by taking the residues given by the poles
from the propagators. By the usual trick we have to take into account the exponential 
factors in the matrix element to argue in which way to close the integration contour and 
thereby we introduce an ordering of the light cone coordinates by
\begin{eqnarray}
  \Theta^{(a)} &=& \Theta(y_1^--y_3^-) \Theta(-y_4^-) \> , \\
  \Theta^{(b)} &=& \Theta(-y_3^-) \Theta(y_3^--y_4^-) \> , \\ 
  \Theta^{(c)} &=& \Theta(y_1^--y_4^-) \Theta(y_4^--y_3^-)
\end{eqnarray}
for diagrams (a), (b) and (c) respectively.

At this stage we see that the limits $r_1^\perp, {r'}_1^\perp \rightarrow 0$,
are trivial in the sense that there are no poles or zeros appearing by taking this limit.
We can therefore set these transverse momenta to zero. This implies
$K^{\prime\perp}=K^\perp$ and $x'=x=-{K^\perp}^2 / \xi_2 S$.
Now we add up the contributions of all three graphs. Note that the pole
integrations introduce minus signs for graphs (2) and (3):
\begin{widetext}
\begin{multline}
  \label{eq:sep}
  \Theta^{(a)} H^{(a)}_{\mu\nu}\Gamma^{(1)} 
  - \Theta^{(b)} H^{(b)}_{\mu\nu}\Gamma^{(2)}
  - \Theta^{(c)} H^{(c)}_{\mu\nu}\Gamma^{(3)} = \\
  \quad = \frac{x \xi_1}{\xi_2 S} \Big[ 2 x P_{1\mu}P_{1\nu} + ( P_{1\mu}K_\nu^\perp
    + P_{1\nu} K_\mu^\perp) \Big] \Theta^{(a)}\Gamma^{(a)} \\
    \qquad- \frac{x \xi_1}{S} \big[ \frac{1}{2} S g_{\mu\nu} - 
    P_{1\mu}P_{2\nu} - P_{1\nu}P_{2\mu} \big] \Theta^{(a)}
    \big[ \Gamma^{(a)}-\Gamma^{(b)}] \big] \\
    - \frac{x\xi_1}{S} \big[ \frac{1}{2} S g_{\mu\nu} - 
    P_{1\mu}P_{2\nu} - P_{1\nu}P_{2\mu} \big] \Gamma^{(b)} 
    \big[ \Theta^{(a)}-\Theta^{(b)}- \Theta^{(c)} \big].
\end{multline}
\end{widetext}

Let us now discuss electromagnetic gauge invariance. After taking the poles all quark lines
attaching to an arbitrary quark photon vertex in either diagram are on the mass shell. 
Therefore electromagnetic gauge invariance at this stage is ensured for each diagram 
separately by simple equations of motion.
Let us denote the three terms  on the three different lines on the right hand side of 
Eq.~(\ref{eq:sep}) by A, B and C starting from above. It is easy to check that C is separately 
gauge invariant and so is the sum of A and B. B and A alone are not gauge invariant as the 
contraction of A with $q^\mu$ gives $1/2 \> x \xi_1 K_\perp^\nu \Theta^{(a)}\Gamma^{(a)}$ 
which is then canceled by the same 
term with different sign originating from contraction of $q^\mu$ with B. As already discussed 
the $\delta$-functions determine $q^\mu = (\xi_1+x) P_1^\mu + \xi_2 P_2^\mu + K^{\perp\mu}$ 
(for $\Gamma^{(a)}$) and $q^\mu = \xi_1 P_1^\mu + \xi_2 P_2^\mu$ (for $\Gamma^{(b)}$).

We will now drop the term C because the term $\Theta^{(a)}-\Theta^{(b)}- \Theta^{(c)}$
requires the partons from the nucleus to be bound together and leads
to a twist-4 contribution which is not nuclear enhanced \cite{Guo:98ht,FSSM:00}. 
The remaining term A+B is only proportional to 
$\Theta^{(a)}$ and is manifestly gauge invariant.
We absorb $\Theta^{(a)}$ into the matrix element
\begin{widetext}
\begin{equation}  
  T_{qg}(\xi_1,x) =
  \int \frac{d{y_1^-}}{2\pi}  
  \frac{d{y_3^-}}{2\pi} 
  d{y_4^-}
  e^{-i\xi_1 P_1^+ y_1^-} e^{-i x P_1^+ (y_3^--y_4^-)} \Theta^{(a)}
  \frac{1}{2} \bra{P_1} \bar q (0) \gamma^+ q(y_1^-)  
  F^{+\omega}(y_3^-) F_\omega^{\phantom{\omega}+}(y_4^-) \ket{P_1}.
\end{equation}
\end{widetext}

We can write the hadronic tensor after summing over all diagrams and 
integrating over $\xi_1$ and $y$ as
\begin{equation}
  W_{\mu\nu} = \frac{128 \pi^5 \alpha_s}{N_c^2} \int_B^1 
  \frac{d{\xi_2}}{\xi_2} f_{\bar q}(\xi_2)  \lim_{K^\perp
  \rightarrow 0} T_{qg}(\xi_1,x)  H_{\mu\nu}
\end{equation}
where $B=Q^2/S$, $\xi_1= Q^2 / (\xi_2 S)$ and 
\begin{multline}
  \label{eq:emhadr}
  H_{\mu\nu} = \Big[ 2 P_{1\mu}P_{1\nu} + \frac{1}{x} 
    ( P_{1\mu}K_\nu^\perp
    + P_{1\nu} K_\mu^\perp) \Big]  \frac{\delta(q_\perp^2 - k_\perp^2)}{(\xi_2 S)^2} \\
  -  \big[ \frac{1}{2} S g_{\mu\nu} -
    P_{1\mu}P_{2\nu} - P_{1\nu}P_{2\mu} \big]   \\ \times
  \frac{\big[ \delta(q_\perp^2 - k_\perp^2) - \delta(q_\perp^2) \big]}{x \xi_2 S^2}. 
\end{multline}
We have introduced $k_\perp^2 = - {K^\perp}^2 \ge 0$.

We note that this expression still has poles due to our choice of 
gauge for the gluon fields, but as checked above electromagnetic gauge
invariance is manifest. It is once more interesting to note that the term 
in above equation which is proportional to the transverse gluon momentum
is needed to ensure gauge invariance for the entire hadronic tensor. This
now justifies the careful treatment of the collinear limit.

We are ready to contract the hadronic tensor with
the photon polarization tensor. Since $q^\mu H_{\mu\nu}$ is zero this is simply a
contraction with $-g^{\mu\nu}$.
The result is simple and finally allows to take the limit 
$K^\perp \to 0$ unambiguously, leading to  
\begin{equation}
  H_{\mu\nu}P^{\mu\nu} = \lim_{k_\perp^2 \rightarrow 0} 
  \frac{1}{k_\perp^2} \big[ \delta(q_\perp^2 - k_\perp^2) - 
  \delta(q_\perp^2) \big] = - \delta' (q_\perp^2)
  \, .
\end{equation}
The final result for the cross section is
\begin{multline}
  \frac{d \sigma_1}{d Q^2 d q_\perp^2} = \sum_q \frac{4\pi c_q^2 \alpha^2}{3 N_c Q^2 S} 
  \\ \times \int_B^1 \frac{d \xi_2}{\xi_2} T_{qg}(\xi_1) f_{\bar q}(\xi_2) 
  \left[ - \frac{4\pi^2 \alpha_s}{N_c} \delta'(q_\perp^2) \right] 
\end{multline}
with the known soft hard matrix element $T_{qg}(\xi_1) = T_{qg}(\xi_1,x=0)$.
A second term proportional to $T_{\bar q g} f_q$ has to be added but is not explicitly shown.
This indeed coincides with the result in \cite{Guo:98jb}. We introduced a subscript ``1'' with
the cross section to indicate that this is the contribution with one pair of gluon
fields in the matrix element.

\subsection{Covariant gauge}

Let us briefly go through the corresponding calculation in covariant gauge, since this
will be the technique we use later for arbitrary twist.
We can again start from Eq.~(\ref{eq:hadtensor}) and proceed
along the same way as before, keeping transverse momenta for the gluon fields. However in
covariant gauge we use
\begin{equation}
  P_1^+ A^\rho P_1^+ A^\sigma \approx P_1^\rho A^+ P_1^\sigma A^+ .
\end{equation} 
We therefore define
\begin{equation}
  H_{\mu\nu}^{(i)} = (\xi_2 S)^{-2} P_1^\rho P_1^\sigma H_{\mu\nu,\rho\sigma}^{(i)},
\end{equation}
where $H_{\mu\nu,\rho\sigma}^{(i)}$ are the traces
for the graphs $i=a$, $b$ and $c$. In covariant gauge we have to carry out the 
collinear expansion to the second order in transverse momentum, applying the operator 
\begin{equation}
  \frac{1}{2!} K_\lambda^\perp K_\kappa^\perp \frac{d^2}{d K_\lambda^\perp 
  d K_\kappa^\perp}\Big|_{K^\perp\to 0}
\end{equation} 
to the hard part. The explicit factors of transverse momentum are used to obtain 
field strength tensors in the matrix element via
\begin{multline}
  \label{eq:gluoncov}
  K^\perp_\lambda A^+ (y_3) K^\perp_\kappa A^+ (y_4) \\ \approx  
  F^{+\omega}(y_3) F_\omega^{\phantom{\omega}+}(y_4) 
  \left( - g^\perp_{\lambda\kappa} /2 \right) .
\end{multline}

After performing the pole integrations in $x$ and 
$x'$ and integrating
over $\xi_1$ and $y$ the hadronic tensor for diagram (a) is
\begin{multline}
  \label{eq:hadtensor3}
  W_{\mu\nu}^{(a)} = \frac{128 \pi^5 \alpha_s}{N_c^2} \int\frac{d \xi_2}{\xi_2} 
  f_{\bar q}(\xi_2)
  \frac{1}{\xi_1}  \left( - \frac{ g^\perp_{\lambda\kappa}}{2} \right) \\ 
  \frac{1}{2!} \frac{d^2}{d K^\perp_\lambda d K^\perp_\kappa}\Big|_{K^\perp\to 0} 
  T_{qg}(\xi_1,x) \, {H}_{\mu\nu}^{(a)} \, \delta(q_\perp^2+K^{\perp 2})
\end{multline}
where now $x=x'=k_\perp^2/\xi_2 S$ and $\xi_1 = Q^2 / \xi_2 S$.
Since 
\begin{equation}
  -g^{\mu\nu} H_{\mu\nu}^{(a)} = \xi_1 \xi_2 S,
\end{equation}
the derivatives can only act on the transverse $\delta$-function and the phases in the matrix element.

For diagrams (b) and (c) Eq.~(\ref{eq:hadtensor3}) holds with an additional ``$-$'' sign and
modified $\Theta$-functions $\Theta^{(b)}$ and $\Theta^{(c)}$. Furthermore the transverse 
$\delta$-function is just $\delta(q_\perp^2)$ and
$-g^{\mu\nu} H_{\mu\nu}^{(b)} = -g^{\mu\nu} H_{\mu\nu}^{(c)} = \xi_1 \xi_2 S$.
Therefore from the asymmetric diagrams (b) and (c) we can only get nuclear enhanced
contributions with derivatives on the phases in the matrix element. 
When we sum over all diagrams the terms with derivatives on phases are again proportional to
\begin{equation}
  \Theta^{(a} - \Theta^{(b)} - \Theta^{(c)} \to 0
\end{equation}
and therefore show no nuclear enhancement --- confer the discussion below Eq.~(\ref{eq:sep}). 
We conclude that the only nuclear enhanced contribution can arise 
from the transverse $\delta$-function of the symmetric diagram. We have 
\begin{multline}
  \left( - \frac{ g^\perp_{\lambda\kappa}}{2} \right)
  \frac{1}{2!} \frac{d^2}{d K^\perp_\lambda d K^\perp_\kappa}\Big|_{K^\perp\to 0}
  \delta(q_\perp^2+{K^\perp}^2) \\ = - \delta'(q_\perp^2).
\end{multline}
This implies
\begin{multline}
  \frac{d \sigma_1}{d Q^2 d q_\perp^2} = \frac{4\pi c_q^2 \alpha^2}{3 N_c Q^2 S} \\ \times
  \int_B^1 \frac{d \xi_2}{\xi_2} T_{qg}(\xi_1) f_{\bar q}(\xi_2) 
  \left[ - \frac{4\pi^2 \alpha_s}{N_c} \delta'(q_\perp^2) \right]. 
\end{multline}
which coincides with the result in light cone gauge.

Let us shortly discuss the result of the twist-4 calculation.
$\delta$-functions and their derivatives, strictly speaking, are only defined when they are 
integrated. This applies also to our result. We have to integrate over the transverse 
momentum in order to obtain an observable quantity. 
Let us define moments $Q_\perp^{2m}$ of the transverse momentum spectrum by
\begin{equation}
  Q_\perp^{2m} = \int_0^\infty d q_\perp^2 \> q_\perp^{2m} \frac{d\sigma}{d Q^2 d q_\perp^2}.
\end{equation}
We recall the leading twist result for the cross section in leading order in $\alpha_s$
\begin{equation}
 \frac{d \sigma_0}{d Q^2 d q_\perp^2} = \frac{4\pi c_q^2 \alpha^2}{3 N_c Q^2 S}
  \int_B^1 \frac{d \xi_2}{\xi_2} f_{q}(\xi_1) f_{\bar q}(\xi_2) \delta(q_\perp^2) .
\end{equation}

A simple integration over $q_\perp^2$ 
gives zero which implies that there is no nuclear enhanced twist-4 contribution from these 
graphs to the spectrum integrated over the transverse momentum
\begin{equation}
  (Q_\perp^0)_1  = \frac{d \sigma_1}{d Q^2} = 0. 
\end{equation}
The subscript ``1'' on the left hand side refers to the contribution from $\sigma_1$.
This is in disagreement with the results of \cite{QZ}.
However, when we look at the second moment of the transverse momentum spectrum, we get a nuclear
enhanced twist-4 contribution \cite{Guo:98jb}
\begin{equation}
  (Q_\perp^2)_1 =  
  \frac{4\pi c_q^2 \alpha^2}{3 N_c Q^2 S} 
  \int_B^1 \frac{d \xi_2}{\xi_2} T_{qg}(\xi_1) f_{\bar q}(\xi_2)  \frac{4\pi^2 \alpha_s}{N_c} .
\end{equation}
From leading twist we obtain
\begin{align}
  (Q_\perp^0)_0 &= \frac{4\pi c_q^2 \alpha^2}{3 N_c Q^2 S}
  \int_B^1 \frac{d \xi_2}{\xi_2} f_{q}(\xi_1) f_{\bar q}(\xi_2) ,   \\
  (Q_\perp^2)_0 &= 0
\end{align}
for the first two moments. All higher moments vanish both for twist-2 and twist-4.

The soft-hard matrix element is usually approximated by
\begin{equation}
  T_{qg}(\xi_1) \approx \lambda^2 A^{4/3} f_q(\xi_1) 
\end{equation}
where $f_q$ is the nuclear parton distribution for a quark normalized to one nucleon and 
$\lambda^2 = 0.01$ GeV$^2$ \cite{Guo:98ht,Guo:98jb,FSSM:00}.
This leads to an additional broadening of the transverse momentum spectrum in $p+A$ collisions 
compared to $p+p$ collisions 
\begin{equation}
  \Delta \langle q_\perp^2 \rangle = \frac{ [ Q_\perp^2 ]_{pA}}{
  [ Q_\perp^0 ]_{pA}} - \frac{[ Q_\perp^2 ]_{pp}}{[ Q_\perp^0 ]_{pp}} .
\end{equation}
To twist-4 accuracy we have $Q_\perp^{2m} = (Q_\perp^{2m})_0 + (Q_\perp^{2m})_1$.
Using the model for the soft-hard matrix element we obtain \cite{Guo:98jb}
\begin{equation}
  \Delta \langle q_\perp^2 \rangle 
  = \frac{4\pi^2 \alpha_s}{N_c} \lambda^2 A^{1/3}.
\end{equation}
The transverse momentum broadening with the characteristic $A^{1/3}$ scaling was observed by 
the Fermilab E772 collaboration \cite{McGMP:99}.

\section{Beyond double scattering}

\subsection{Preliminaries}

\begin{figure}
  \begin{center}
  \epsfig{file=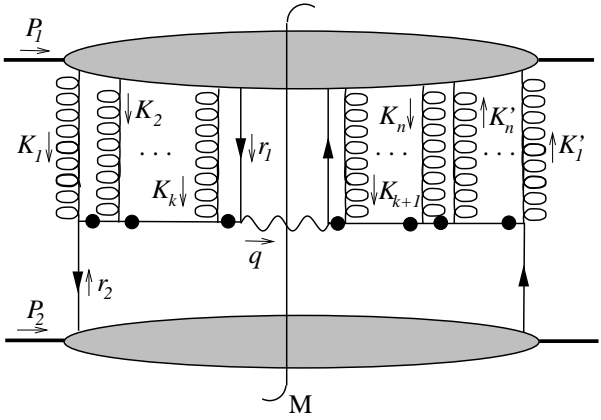,width=8cm}\hfil
  \end{center}
  \caption{Scattering off 2n gluons. In this example $k$ gluons ($k<2n$) are on the left side
   and $2n-k$ gluons on the right side of the cut. The gluons are ordered to pairs with 
   momenta $K_i$ and $K_i'$ for one pair. Big dots indicate soft gluon poles.}
  \label{fig:multsc}
\end{figure}

\begin{figure}[tb]
  \begin{center}
  \epsfig{file=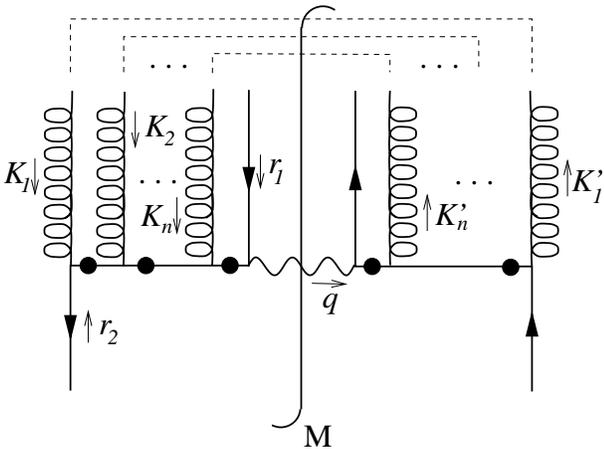,width=8cm}\hfil
  \end{center}
  \caption{The symmetric diagram with $n$ gluons on each side, where opposite gluons constitute
  pairs of color singlets, is the only planar diagram which contributes to maximal 
  nuclear enhancement.
  Dotted lines show the planar flow of color through the matrix element.}
  \label{fig:planar}
\end{figure}

The result of the last section raises the question whether terms of even higher twist 
contribute to higher moments of the transverse momentum spectrum. This is indeed the case and 
will eventually allow us to resum the contributions of different twist.
To start we want to pin down what we are looking for. 
We are interested in contributions of Twist-$(2n+2)$, $n\in \mathbb{N}$, 
with maximal nuclear enhancement $A^{n/3}$.
Therefore we take into account operators with $n$ pairs of soft gluons, 
each single pair a color singlet with no flow of momentum between these pairs.
These are necessary conditions to achieve maximal nuclear enhancement.
The diagram in Fig.~\ref{fig:multsc} is an example with $2n$ gluons. 
Furthermore we use the argument, that in a large $N_c$ expansion the leading contribution
is coming from planar diagrams \cite{tHo:74}.
In order to handle the large number of possibilities to group $2n$ gluons together to pairs
we stick to the leading order in $1/N_c$ and only take into account planar diagrams, i.e.\ with
no gluon lines intersecting each other. See Fig.~\ref{fig:planar} for an example

We have seen that the case of twist-4 was simpler to handle in covariant gauge. Therefore 
we will use this gauge here. This requires us to take a derivative of the hard part of order $2n$ with 
respect to the transverse momenta. We denote the gluon momenta with $K_1$, $K_2$, \ldots\ and $K'_1$, 
$K'_2$, \ldots, where always $K_i$ and $K'_i$ belong to one pair. We introduce parton momentum fractions
by 
\begin{equation}
  K_i = x_i P_1 + K^{\perp}_i,  \qquad  K_i^{\prime} = x_i^{\prime} P_1 + K^{\prime\perp}_i.
\end{equation}
The differential operator we have to apply is of the form
\begin{equation}
  \label{eq:multop}
  \frac{1}{(2n)!} \prod_i K_i^{\perp \lambda_i} K_i^{\prime \perp \kappa_i} 
  \frac{d^{2n}}{\prod_i d K_i^{\perp \lambda_i} d K_i^{\prime \perp \kappa_i} }
  \Big|_{K_i^{\prime\perp},K_i^{\perp} \to 0}.
\end{equation}
In order to obtain the matrix element with $2n$ field strength tensors in covariant gauge, we have to 
apply derivatives to all $2n$ momenta once. To keep it simple we investigate the contribution with 
a maximal number of derivatives on the transverse $\delta$-function. This will give us a maximal
broadening effect.

E.g.\ the diagram in Fig.~\ref{fig:multsc} provides a term 
$\delta(q_\perp^2+(K_1^\perp + \ldots + K_k^\perp)^2)$. 
When we keep our requirement in mind that there should be no flow of momentum between the pairs of
gluons we have $K_i^\perp = K_i^{\prime \perp}$. Then application of the operator (\ref{eq:multop})
to this transverse $\delta$-function gives
\begin{widetext}
\begin{equation}
  \frac{1}{(2n)!} 
  \frac{d^{2n}}{\prod_i d K_i^{\perp \lambda_i} d K_i^{\perp \kappa_i} }
  \Big|_{K_i^{\prime\perp},K_i^{\perp} \to 0}
  \delta(q_\perp^2+(K_1^\perp + \cdots + K_k^\perp)^2)
  = \delta_{nk} \frac{2^n}{(2n)!} \delta^{(n)}(q_\perp^2) \prod_i g^{\lambda_i \kappa_i}
\end{equation}
\end{widetext}
where $\delta^{(n)}$ denotes a $n$th derivative of a $\delta$-function.

In the complete Taylor expansion of order $2n$ there are $(2n)! / 2^n$ derivative terms of 
the kind of Eq.~(\ref{eq:multop}) which have to be summed. This exactly cancels the 
combinatorial factor in above equation. We note that we only get a non-vanishing contribution 
if $k=n$, i.e.\ if the diagram is symmetric.
It is straightforward to show that for fixed $n$ the symmetric diagram in 
Fig.~\ref{fig:planar} is the only planar diagram with $2n$ gluons which contributes with 
the maximal number of transverse derivatives. Therefore with the restrictions we have chosen 
we have only one diagram to calculate in covariant gauge for each $n$.

\subsection{Calculation of arbitrary twist}

The hadronic tensor for the diagram with $n$ gluon pairs can be written as
\begin{widetext}
\begin{multline}
  \label{eq:hadtensor4}
  W_{\mu\nu}^{n} = (4\pi\alpha_s)^n C_n
  \int\frac{d \xi_1}{\xi_1} \frac{d \xi_2}{\xi_2} \int \prod_i d x_i d x'_i d^2 K^\perp_i 
  d^2 K^{\prime \perp}_i \> f_{\bar q}(\xi_2)
  {H}_{\mu\nu}^{n} (32\pi^3)\Gamma^n {\cal P}^n
  \\ \times
  \int \frac{d y_0^-}{2\pi} e^{-i\xi_1 P_1^+ y_0^-} 
  \prod_i \int \frac{d{y_i^-}}{2\pi} \frac{d y_i^{\prime -}}{2\pi}
  e^{-i\xi_i P_1^+ y_i^-} e^{i x'_i P_1^+ y_i^{\prime -} }
  \frac{1}{2} \bra{P_1} \bar q (0) \gamma^+ q(y_0^-) \prod_i A^+(y_i^-) A^+(y_i^{\prime -})
  \ket{P_1}.
\end{multline}
\end{widetext}
We start with the evaluation of the color factor for every diagram which is 
\begin{equation}
  C_n = \frac{1}{N_c} \left( \frac{1}{2 N_c} \right)^n.
\end{equation}
For the hard part we get the simple result
\begin{widetext}
\begin{multline}
  - g^{\mu\nu} H^n_{\mu\nu}
   = (\xi_2 S)^{-n} \frac{1}{4} \tr \bigg[ \gamma_\nu \left(-\sla{r}_2 - \sum_{i=1}^n \Sla{K}'_i \right) 
  \gamma_{\sigma_n} \left(-\sla{r}_2 - \sum_{i=1}^{n-1} \Sla{K}'_i \right) \gamma_{\sigma_{n-1}} 
   \cdots \\
   \cdots \gamma_{\sigma_1} \sla{r}_2 \gamma_{\rho_1} \cdots \gamma_\mu \sla{r}_1 \bigg] (- g^{\mu\nu})
   \prod_i P_1^{\sigma_i} P_1^{\rho_i} = \xi_1 \xi_2 S.
\end{multline}
\end{widetext}
This can be proved by repeatedly applying the identity
\begin{equation}
  \Sla{P}_1 (\ldots) \Sla{P}_1 = 2 P_1 \cdot (\ldots) \> \Sla{P}_1 = \xi_2 S \Sla{P}_1
\end{equation}
where $(\ldots)$ stand for one of the parentheses in the equation above. 

The $2n$ propagators provide poles
\begin{multline}
  {\cal P}^n = 
  \frac{1}{x_1 +\frac{(K_1^\perp)^2}{\xi_2 S}+i\epsilon} \> \frac{1}{x_1 + x_2 
  +\frac{(K_1^{\perp 2} + K_2^{\perp 2} )^2}{\xi_2 S} +i\epsilon} 
  \\ \cdots \frac{1}{x'_1 +\frac{(K_1^{\prime \perp})^2 
  - i\epsilon}{\xi_2 S}}
\end{multline} 
and we also have $\delta$-functions
\begin{multline}
  \Gamma^n = \delta \left( Q^2 - \Big(\xi_1 +\sum_i x_i\Big) \xi_2 S - 
  \Big(\sum_i K_i^\perp\Big)^2
  \right) \\ \times
  \delta \left( q_\perp^2 + \Big(\sum_i K_i^\perp\Big)^2 \right) \\ \times
  \delta \left(y - \frac{1}{2} \ln \frac{\xi_1 +\sum_i x_i}{\xi_2} \right).
\end{multline}

After integrating the momentum fractions $x_i$ and $x'_i$ by taking the poles and integrating over 
$y$ and $\xi_1$
we apply the operators from Eq.~(\ref{eq:multop}) to the transverse $\delta$-functions and
use the explicit factors of $K_i^\perp$ in front of the operator together with Eq.~(\ref{eq:gluoncov}).
We obtain a hadronic tensor
\begin{multline}
  \label{eq:hadtensor5}
  - g^{\mu\nu} W_{\mu\nu}^{n} = 
  \frac{32 \pi^3}{N_c} \left( - \frac{8\pi^2 \alpha_s}{2 N_c} \right)^n \\ \times
  \int_B^1 \frac{d \xi_2}{\xi_2} f_{\bar q}(\xi_2) T_{qg^n} (\xi_1) \delta^{(n)}(q_\perp^2)
\end{multline}
where $\xi_1 = Q^2 / \xi_2 S$ and the matrix element with $n$ pairs of gluon operators is 
defined as
\begin{multline}
  \label{eq:multmatrix}
  T_{qg^n} (\xi) =   \int \frac{d{y_0^-}}{2\pi} \int \prod_i \frac{d{y_i^-} 
   d{y_i^{\prime -}}}{2\pi} e^{-i\xi P_1^+ y_0^-} \Theta^n \\  \times
  \frac{1}{2} \bra{P_1} \bar q (0) \gamma^+ q(y_0^-) 
  \\ \times \prod_i F^{+\omega}(y_i^-) F_\omega^{
  \phantom{\omega} +}  (y_i^{\prime -}) q(y_0^-)
  \ket{P_1}.
\end{multline}
The $\Theta$-functions from the pole integrations reflect causality and give an ordering of the 
vertices along the quark lines on the light cone
\begin{multline}
  \Theta^n = \Theta(y_0^- - y_n^-) \Theta(y_n^- - y_{n-1}^-) \cdots \Theta(y_2^- - y_{1}^-)
  \\ \times \Theta(- y_n^{\prime -}) \Theta(y_n^{\prime -} - y_{n-1}^{\prime -}) \cdots 
  \Theta(y_2^{\prime -} - y_{1}^{\prime -}).
\end{multline}

The cross section can finally be written as
\begin{multline}
  \label{eq:multcross}
  \frac{d \sigma_n}{d Q^2 d q_\perp^2} = \frac{4\pi c_q^2 \alpha^2}{3 N_c S Q^2} 
  \left[ - \frac{4\pi^2 \alpha_s}{N_c} \frac{d}{d q_\perp^2} \right]^n \delta(q_\perp^2)  
  \\ \times
  \int_B^1 \frac{d \xi_2}{\xi_2} T_{qg^n}(\xi_1) f_{\bar q}(\xi_2) ,
\end{multline}
including the known cases for $n=0$, 1.
Again we can obtain physical results only if we integrate over the transverse momentum and resolve the 
$\delta$-functions.
From matrix elements with $n$ gluon pairs we get contributions to the $2m$th moment of the 
transverse momentum spectrum
\begin{multline}
   (Q_\perp^{2m})_n  = \delta_{nm} \frac{4\pi c_q^2 \alpha^2}{3 N_c S Q^2}
   \left[ \frac{4\pi^2 \alpha_s}{N_c} \right]^m \\ \times
   \int_B^1 \frac{d \xi_2}{\xi_2} T_{qg^n}(\xi_1) f_{\bar q}(\xi_2) .
\end{multline}

When applying the derivatives in the differential operator (\ref{eq:multop}) to the hard part
we have chosen to take all derivatives on the transverse $\delta$-function. We know that in the
twist-4 case ($n=1$) this was the only way to obtain nuclear enhancement. This is no longer 
guaranteed for $n>1$. The derivatives can also act on the phases in the matrix element. 
Moreover more graphs and orderings of gluon pairs can contribute. This has to be taken into 
account for a more quantitative analysis. 
In general, from a diagram with $n$ gluon pairs we expect a cross section which can be
written as a sum
\begin{equation}
  \label{eq:fullsum}
  \frac{d\sigma}{d Q^2 d q_\perp^2} = \sum_{k=0}^n \frac{V_k}{Q^{2(n-k)}} 
  \delta^{(k)}(q_\perp^2) .
\end{equation}
Lower derivatives in $q_\perp^2$ are compensated by explicit factors of $1/Q^2$.
The $V_k$ are coefficients which are unknown at present except for the case $k=n$ where
they can be taken from Eq.~(\ref{eq:multcross}).

Therefore from a diagram with $n$ gluon pairs we would get 
terms contributing to all momenta of 
the spectrum lower than $n$. We were already arguing that our result, where we only keep 
the terms $k=n$, will give us a maximal broadening effect. We will discuss later how 
this shows up after summation of all twists and how the subleading terms with less derivatives 
will alter the result.

\subsection{Resummation}

In order to sum contributions of arbitrary twist, one has to establish relations between the 
matrix elements $T_{qg^n}$ of different twist. Our approach will be to develop a simple 
model for these matrix elements. The usual way to treat the twist-4 soft hard matrix element 
is to factorize the soft gluons from the quarks. For that one inserts complete sets of nuclear
states
\begin{equation}
  1 = \sum_\alpha \int \frac{dp}{N} \ket{p;\alpha} \bra{p;\alpha}, 
\end{equation}
between the pairs of parton operators.
$N$ is a normalization factor. $\alpha$ is a symbolic index for all the excited states 
of the nuclear system and $p$ is the momentum normalized to one nucleon.
Later we approximate sum and integral by the contribution of the 
lowest state with the given momentum $\ket{P_1}$. We assume that this is a good approximation
when there is no momentum transfer between the parton pairs. 

Furthermore we introduce relative and absolute coordinates for the gluon pairs
\begin{eqnarray}
  z_i &=& (y_i^- - y_i^{\prime -})/2, \\
  Z_i &=& y_i^- + y_i^{\prime -}.
\end{eqnarray}
Since gluon pairs are bound to color singlets we can assume that $|z_i| \ll R_A$ for large
$A$, where $R_A$ 
is the nuclear radius.
This enables us to use
\begin{equation}
  \Theta^n \approx \Theta(-Z_n) \Theta(Z_n-Z_{n-1}) \cdots \Theta(Z_2-Z_{1}).
\end{equation}
Corrections to that will lead to contributions which are suppressed by powers of $A^{1/3}$ 
compared to the full nuclear enhancement.

We introduce bilocal gluon correlators
\begin{equation}
  G(Z_i) = \int \frac{dz_i}{2\pi N} \bra{P_1} F^{+\omega}(Z_i-z_i) F_\omega^{
  \phantom{\omega} +}  (Z_i+z_i) \ket{P_1}.
\end{equation}
We assume now that the value of this correlator does not depend strongly on the absolute position
of the gluon pair inside the nucleus, so that $G(Z_i)\approx G_0$. 
Our assumptions lead to 
\begin{multline}
  T_{qg^n} (\xi) = A f_q(\xi) G_0^n \int \prod_i d Z_i \\ \times
  \Theta(-Z_n) \Theta(Z_n-Z_{n-1}) 
  \cdots \Theta(Z_2-Z_{1}) \\ =
  A f_q(\xi) \frac{1}{n!} \prod_i \left( G_0 \int_{-R_A}^0 d Z_i \right).
\end{multline}
The term in parentheses scales with $A^{1/3}$ and we set it equal to $\lambda^2 A^{1/3}$. 
$\lambda^2$ was already introduced for modeling the twist-4 matrix element.
Finally we have
\begin{equation}
  T_{qg^n} (\xi) = \frac{1}{n!} A^{1+n/3} \lambda^{2n} f_q(\xi).
\end{equation}
As expected the twist-(2n+2) matrix element scales with $n$ additional powers 
of $A^{1/3}$.

The assumptions we have used can be summarized by the statement that all QCD dynamics is 
reduced to the QCD substructure of the single nucleons in the nucleus, whereas the nuclear 
effects are taken to be purely geometrical in nature. This might be quite reasonable to estimate 
leading effects. We remember that in the case of parton distributions the idea
of a superposition of parton distributions of the individual nucleons is also leading to 
a reasonable first guess for the nuclear parton distributions. Dynamical nuclear effects 
then enter as corrections which might be as large as 20 or 30\%, showing up as phenomena like 
shadowing and the EMC effect \cite{shadowing}.  

When we plug our model into Eq.~(\ref{eq:multcross}), we get
\begin{multline}
  \label{eq:multmodel1}
  \frac{d \sigma_n}{d Q^2 d q_\perp^2} = \frac{4\pi c_q^2 \alpha^2}{3 N_c S Q^2} 
  \int_B^1 \frac{d \xi_2}{\xi_2} f_{q}(\xi_1) f_{\bar q}(\xi_2) \\ \times \frac{1}{n!}
  \left[ - \frac{4\pi^2 \alpha_s}{N_c}
  A^{1/3} \lambda^2 \frac{d}{d q_\perp^2} \right]^n \delta(q_\perp^2) .
\end{multline}
Summing up all twists leaves us with a shift operator, acting on the $\delta$-function
\begin{multline}
  \label{eq:multmodel2}
  \frac{d \sigma}{d Q^2 d q_\perp^2} = \sum_n \frac{d \sigma_n}{d Q^2 d q_\perp^2} \\ =
  \exp\left[ - \frac{4\pi^2 \alpha_s}{N_c}
  A^{1/3} \lambda^2 \frac{d}{d q_\perp^2} \right] \frac{d \sigma_0}{d Q^2 dq_\perp^2} \\ =
  \frac{d \sigma_0}{d Q^2} \delta\left( q_\perp^2 - \frac{4\pi^2 \alpha_s}{N_c}
  A^{1/3} \lambda^2 \right)
\end{multline}

At first glance this is a surprising result. Let us note that the summation of an infinite 
series of derivatives of $\delta$-functions can give a well-defined smooth function. 
That our particular result is again of $\delta$ shape is in some sense accidental and due to 
our simple model. 

However we remember that we already expected a result which gives us a maximal
nuclear broadening. This was because we dropped all terms with less than $n$ transverse
derivatives in the general expression in Eq.~(\ref{eq:fullsum}). The result of a 
shifted $\delta$-function, where there is no contribution with $q_\perp=0$ left,
is the outcome of cutting the sum in (\ref{eq:fullsum}). We expect contributions with 
less derivatives on the transverse $\delta$-function to alter the shape
through explicit $1/Q^2$ dependence which is absent in our result. 
This should lead to contributions which are less shifted and therefore, when added, 
filling the ``valley'' between $q_\perp=0$ and the maximal shift value of 
$q_\perp^2 = 4\pi^2 \alpha_s A^{1/3} \lambda^2 / N_c$.

A more rigorous argument would be that
for a given $n$, because of the lower number of derivatives in the terms 
dropped in Eq.~(\ref{eq:fullsum}), these add to lower moments,  
therefore favouring lower values of $q_\perp^2$ in the function reconstructed from these
moments. A more quantitative analysis is needed here. We shall report on that in a forthcoming
publication.
However our result allows to give the order of the smearing of transverse momentum through
nuclear effects. It is of the order of
\begin{equation}
  \bar q_\perp^2  = 4\pi^2 \alpha_s A^{1/3} \lambda^2 / N_c \approx (400\text{ MeV})^2
\end{equation}
for large nuclei and for $\lambda^2 = 0.01\text{ GeV}^2$. 
 
We have to keep in mind that we are dealing only with additional nuclear effects that
add to the smooth transverse momentum spectrum in $p+p$ collisions and which can be described 
theoretically by higher order calculations and resummation of radiative corrections. 
A resummation of nuclear higher twist is an additional effect, though not completely
independent from a resummation of radiative corrections. The scale of these nuclear
effects is determined by $\bar q_\perp$. This was already claimed to be the scale 
some time ago based on pure twist-4 calculations \cite{LQS:94sv}, but is now confirmed to hold
also if one takes into account matrix elements of arbitrary twist.
$\bar q_\perp$ has also been connected to the work of Baier, Dokshitzer,
Muller, Peigne and Schiff on nuclear momentum broadening \cite{BDMPS}. In this work
also multiple scatterings are effectively summed up.

\section{Conclusions}

We have calculated the transverse momentum broadening of Drell Yan pairs in $p+A$ 
collisions by taking into account nuclear enhanced higher twist effects. We presented 
explicit computations of the twist-4 contribution in light cone gauge and covariant 
gauge, confirming earlier results in covariant gauge. 

Furthermore we obtained results for the maximal broadening effect of leading order
diagrams with arbitrary twist. 
By relating the corresponding higher twist matrix elements through a particular model 
we were able for the first time to give a closed form for the sum over all twist 
contributions. We found that
the effect of nuclear broadening is of the order of $\bar q_\perp \approx 400$ MeV 
for large nuclei.

The exact shape of the broadening effect can be obtained by the study of terms which are
subleading in the number of derivatives in transverse momentum. This has to be investigated 
in the future.

\begin{acknowledgments}
The author is grateful to O.~Teryaev, B.~M\"uller, A.~Sch\"afer, J.~Qiu and G.~Sterman
for useful discussions. The author wants to thank J.\ Raufeisen for pointing out a mistake
in an early version of the manuscript.
The author acknowledges support from the Feodor Lynen program of
the Alexander von Humboldt Foundation. This work was supported in part by 
DOE grant DE-FG02-96ER40945 and BMBF.
\end{acknowledgments}


\newpage

\end{document}